# A setup for studies of photoelectron circular dichroism from chiral molecules in aqueous solution

Sebastian Malerz,[1] Henrik Haak,[1] Florian Trinter,[1,2] Anne B. Stephansen,[1,3] Claudia Kolbeck,[1,4] Marvin Pohl,[1,5] Uwe Hergenhahn,[1] Gerard Meijer,[1] and Bernd Winter[1,*]

[1] *Molecular Physics Department, Fritz-Haber-Institut der Max-Planck-Gesellschaft, Faradayweg 4-6, 14195 Berlin, Germany*

[2] *FS-PETRA-S, Deutsches Elektronensynchrotron (DESY), Notkestraße 85, 22607 Hamburg, Germany*

Current Addresses:

[3] Seaborg Technologies, Titangade 11, 2200 Copenhagen, Denmark

[4] sonUtec GmbH, Mittlere-Motsch-Straße 26, 96515 Sonneberg, Germany

[5] Department of Chemistry, University of California, Berkeley, CA 94720, USA

[*] Corresponding author: *winter@fhi-berlin.mpg.de*

## ABSTRACT

We present a unique experimental design that enables the measurement of photoelectron circular dichroism (PECD) from chiral molecules in aqueous solution. The effect is revealed from the intensity difference of photoelectron emission into a backward-scattering angle relative to the photon propagation direction when ionizing with circularly polarized light of different helicity. This leads to asymmetries (normalized intensity differences) that depend on the handedness of the chiral sample and exceed the ones in conventional dichroic mechanisms by orders of magnitude. The asymmetry is largest for photon energies within several electron volts above the ionization threshold. A primary aim is to explore the effect of hydration on PECD. The modular and flexible design of our experimental setup *EASI* (**E**lectronic structure from **A**queous **S**olutions and **I**nterfaces) also allows for detection of more common photoelectron angular distributions (PADs), requiring distinctively different detection geometries, and typically using linearly polarized light. A microjet is used for liquid-sample delivery. We describe *EASI*'s technical features and present two selected experimental results, one based on synchrotron-light measurements and the other performed in the laboratory, using monochromatized He-II α radiation. The former demonstrates the principal effectiveness of PECD detection, illustrated for prototypic gas-phase fenchone. We also discuss the first data from liquid fenchone. In the second example we present valence photoelectron spectra from liquid water and NaI aqueous solution, here obtained from a planar-surface microjet (flatjet). This new development features a more favorable symmetry for angle-dependent photoelectron measurements.

## I. INTRODUCTION

### A. General considerations

Photoelectron spectroscopy (PES) studies from liquids and particularly from water and aqueous solutions, mostly in conjunction with a liquid microjet,[1, 2] have contributed tremendously to our current understanding of aqueous-phase electronic structure. An experimental focus has been on core-level PES,[2, 3] with far less studies directed at the lowest-ionization energies, although the latter govern chemical reactivity.[4, 5] Core-level spectra, typically measured with tunable soft-X-ray photons from synchrotron radiation beamlines, have identified chemical shifts of solutes, pH-dependent protonation and de-protonation,[6-8] solvent and solute interfacial depth profiles,[9, 10] as well as several non-local electronic relaxation processes, such as intermolecular Coulombic decay (ICD).[11] In most of these studies, electrons with some tens to hundreds of eV kinetic energies (KE) were detected. Single-photon ionization-threshold phenomena in the aqueous phase, corresponding to generation of photoelectrons with kinetic energies typically smaller than 20 eV, have been barely addressed.[12] This is despite their significant relevance, including the increase of photoionization cross sections near an ionization edge, the yet to be demonstrated liquid-phase post-collision interaction (PCI),[13, 14] or the potential presence of photoelectron circular dichroism (PECD)[15-19] in the ionization of liquids. Enabling the exploration of near-threshold ionization phenomena, and particularly aqueous-phase PECD, has been a major motivation to build *EASI* (**E**lectronic structure from **A**queous **S**olutions and **I**nterfaces), a unique, versatile liquid-microjet photoelectron spectroscopy setup.

Our interest in PECD is motivated by the fact that it uniquely connects molecular electronic structure to chirality.[15] The effect manifests as a forward–backward asymmetry in the photoelectron emission intensity from chiral molecules, measured with respect to the propagation direction $\vec{k}$ of circularly polarized light (CPL), the sign of which depends on the helicity of the ionizing radiation (*left* or *right*-handed, *l*-CPL or *r*-CPL). The magnitude of PECD is expressed via the chiral anisotropy parameter $b_1$. Furthermore, the PECD mechanism is solely based on electric dipole transition amplitudes, which leads to much stronger effects than found in conventional circular dichroism methods.[19] Since chirality is a universal property, and of particular importance for biochemically relevant complexes in aqueous solution, it is highly desirable to quantify PECD in an aqueous environment and understand the molecules' possible chiral imprint on their solvation shells.

The principal geometry of a PECD measurement is illustrated in Figure 1a. However, application to the liquid phase requires that several experimental and technical hurdles are overcome, calling for novel and dedicated experimental designs. For gas-phase targets, PECD studies can be readily and very efficiently performed with a velocity map imaging (VMI) spectrometer,[20] which provides high electron collection efficiency by simultaneous and angle-resolved acquisition of the electron signal in all emission directions. However, currently available VMI spectrometers are not compatible with liquid jets for several reasons: (1) A liquid jet represents a dielectric filament of improperly defined charge state, thus introducing undesired electric-field perturbations near the actual ionization region. (2) For sole geometrical reasons, VMI cannot image the full photoelectron angular distribution (PAD) from a cylindrical jet since photoelectrons born inside the solution engage in multiple electron-scattering processes, mostly with water molecules.[12] These electrons may even be directed away from the liquid – vacuum interface into the solution, or if reaching the detector, they will contribute to a signal background that will be difficult to quantify. Using a planar liquid microjet (see Section III. B) might be advantageous since electrons would be detected from a single surface orientation, rather than from a curved surface. Admittedly though, an increased water vapor pressure from a flatjet is likely to result in additional disturbing electron scattering. (3) An additional, more technical complication arises from the considerable background vapor pressure in a liquid-jet experiment which for highly volatile water and aqueous solutions may well be in the ~$10^{-3}$ mbar (for the flatjet) to $10^{-5}$ mbar range. Correspondingly, the successful implementation of the VMI technique with a liquid jet target remains a challenging technical goal. A first approach towards a technical realization has been attempted very recently for non-electrically-conductive solutions. Yet, particularly the consequences of electron scattering in the liquid phase have been barely elaborated on.[21] The same challenges hold for COLTRIMS-type setups, which also have a history of providing important results on gas-phase PECD,[22] but have yet to be implemented with liquid-phase targets.

Here, we take a more conventional, simpler, and currently feasible approach by using a state-of-the-art hemispherical electron analyzer (HEA) equipped with a differentially pumped pre-lens section (capable of near-ambient pressure experiments),[2] and mounted in a geometry compliant with the requirements of PECD detection. Magnetic fields in the region where the liquid jet is ionized are carefully shielded by a full µ-metal encasement, enabling the detection of photoelectrons and elastically and inelastically scattered electrons down to near-zero-eV kinetic energy with quantitative accuracy,[12, 23] as required for studying any (near) ionization-threshold phenomena. This includes the measurement of the low-energy cutoff and low-energy tail in a water or aqueous-solution PE spectrum.[12, 23] Also, in gas-phase studies, PECD was found to be most prominent at electron kinetic energies smaller than approximately 15 eV.[15-19]

A major drawback imposed by the *geometric* constraints in a liquid-jet experiment is that the dichroic effect, resulting in different intensities emitted in forward and backward directions, cannot be directly and simultaneously measured. Instead, the signal intensity, obtained at a (necessarily) fixed detection angle of our hemispherical electron analyzer has to be collected for alternating CPL helicity. A similar detection scheme has been previously used to demonstrate core-level PECD in the gas phase.[24] Yet, extension to liquid-jet PECD (LJ-PECD) experiments entails major technical considerations and developments, which will be detailed below.

A suitable radiation source for our PECD experiments is the synchrotron radiation delivered from a helical undulator (*e.g.,* APPLE-II).[25, 26] But the flexible design of *EASI* also enables PAD measurements to be carried out using linearly polarized synchrotron radiation. For this purpose, *EASI* is devised to detect signals within the plane perpendicular to the propagation of the light ('dipole plane', see Figure 1b) at three alternative, fixed detection angles: 0° (horizontal, in the floor plane, parallel to the polarization vector), 54.7° (magic angle), 90° (perpendicular to the floor and polarization vector). These optional geometries are relevant when only linear horizontal polarization is available, which is the case for many beamlines at synchrotron-light facilities. Photoelectrons from most orbitals are emitted preferentially in the direction of the polarization vector,[27] while electrons from Auger or ICD processes typically feature an isotropic emission pattern.[28] Then, choosing the 0°-geometry for photoelectrons, and 90°-geometry for Auger electrons will yield relatively larger intensities of the respective spectral ranges. The 54.7°-geometry is used to explicitly suppress any angular distribution effects (see below), for instance, when comparing relative signal intensities from ionization of different orbitals for quantitative analysis of relative solute concentration. If linear polarization with a variable orientation of the polarization ellipse is available, any detection angle within 0°-90° can be realized for any of the three geometries, and photoelectron angular distributions (PADs) can be fully mapped out allowing for a determination of the common (dipolar) anisotropy parameter, $\beta$, from aqueous solution, of both water solvent and solute. This parameter can reveal hydrogen-bonding-induced orbital structure changes at the solution–vacuum interface[29] and also provides insight into the molecular structure at such interfaces.[30]

In the following, we will describe the overall design of *EASI* and its components, including the main technical specifications and its principal detection geometries. We close by presenting experimental results to highlight the performance of *EASI*. These include core-level PECD measurements from gas- and liquid-phase fenchone, and regular valence PE spectra obtained from a planar microjet (flatjet) using unpolarized He-II $\alpha$ (40.814 eV) radiation. It is useful though to first review the aforementioned anisotropy parameters, which are relevant for PAD and specifically PECD experiments.

**B. Photoelectron angular distributions in single-photon ionization**

The directional anisotropy of the photoemission process from molecules has played a decisive role in the conceptual design of *EASI*. We therefore review here the main aspects determining PADs. We restrict ourselves to single-photon photoionization of a randomly oriented target within the dipole approximation by light in a pure polarization state $p$, with $p = +1$ designating *l*-CPL in the sense of the optical convention, $p = 0$ linear, and $p = -1$ corresponding to *r*-CPL.[24, 31] The PAD describes differential photoelectron intensities as a function of the angle between a principal symmetry axis and the detection direction. In the case of unpolarized light or CPL, the symmetry axis is the light-propagation direction $\vec{k}$ (Figure 1a), whereas for linearly polarized light, it is the direction of the electric field vector $\vec{E}$ (Figure 1b). In the following, we distinguish these cases by denoting the respective angles as $\theta$ and $\varphi$.

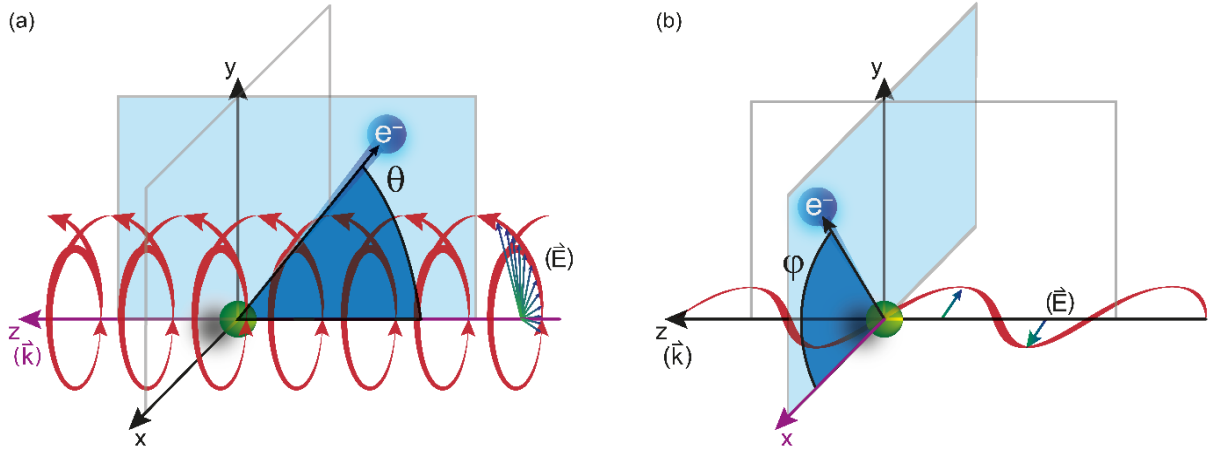

**Figure 1:** Sketch of the relevant principal symmetry axes and respective angles for PES experiments using circularly polarized light (a) or linearly (horizontally) polarized light (b); also see Equations (1)-(3). The green circle indicates the ionization region. In panel (a), the important parameter is the angle $\theta$ spanned between the propagation direction (wave vector, $\vec{k}$) of the circularly polarized light and the electron detection axis, shown here for detection in the direction opposite to $\vec{k}$ (backward-scattering geometry). In panel (b), the important parameter is the angle $\varphi$ spanned between the electric field vector $\vec{E}$ and the electron detection axis, shown here in the plane perpendicular to the floor plane and the propagation direction (dipole plane).

PADs are uniquely connected with several important electronic-structure properties, for instance, photoionization dynamics, based on interfering photoelectron partial waves. Coupling of the electron and photon angular momenta introduces certain symmetry properties and constraints. These symmetry conditions of the experiment determine which terms in the angular distribution function contribute to the PAD. In the following we restrict ourselves to the electric dipole approximation. The influence of magnetic and higher-order electric multipoles on the PADs of linear molecules for single-photon photoionization at photon energies below 1 keV was experimentally found to be small.[32, 33] We expect the same also for chiral-specific non-dipole terms, as derived in Ref. 34. This has been discussed in some detail in the literature.[35] The angular distribution function for perfectly linearly polarized light (LPL) can then be written in the form[36-38]

$$I(\varphi) \propto 1 + \beta P_2(\cos \varphi) \qquad (1),$$

where $P_2(x)$ is the second-order Legendre polynomial which provides the non-isotropic part of the overall distribution, and $\varphi$ is the angle between the linear polarization vector $\vec{E}$ and the direction of photoelectron emission (Figure 1b). The anisotropy parameter $\beta$ is constrained to values $-1 \leq \beta \leq +2$ specifying the magnitude of the emission anisotropy which ranges from a pure $\cos^2(\varphi)$ to a $\sin^2(\varphi)$ form, and therefore possesses mirror symmetry about the principal symmetry axis which is always the polarization vector. For CPL, the PAD is governed by a similar expression (valid only for non-chiral targets, see below):

$$I(\theta) \propto 1 - \frac{\beta}{2} P_2(\cos\theta) \qquad (2),$$

with $\theta$ defined as the angle between the photon propagation vector $\vec{k}$ and the direction of photoelectron emission (see Figure 1a). The second Legendre polynomial has a zero crossing at $x = \cos(54.7°)$. At this particular (magic) angle, the measured differential cross section for any transition will become independent of its $\beta$-value and thus proportional to its total cross section.

Less widely recognized is that these equations are just special (though common) sub-cases of a more general expression:[19, 38]

$$I_p(\theta) \propto 1 + b_1^p P_1(cos\theta) + b_2^p P_2(cos\theta) \qquad (3).$$

The equation is written with the understanding that the variable $\theta$ is replaced by $\varphi$ in the linearly polarized case. The coefficients $b_n^p$ are determined by the photoionization dynamics and depend on the photon polarization state $p$ and the radial dipole amplitudes between the molecular initial and ionized state. For the $P_2(x)$ terms, this leads to the relation $\beta = b_2^0 = -2b_2^{\pm 1}$. Moreover, $b_1^0 = 0$ while $b_1^{\pm 1}$ also vanishes for achiral molecules; in such circumstances, this general expression (3) reduces to the well-known former forms (1) and (2).

Particularly relevant for the present work is that for the specific case of a chiral molecule ionized with CPL, the $P_1(x)$ (first-order Legendre polynomial) coefficients no longer vanish for symmetry reasons.[39] Furthermore, they switch signs with respect to a change of light polarization: $b_1^{+1} = -b_1^{-1}$. The same change in sign of the $b_1^{\pm 1}$ coefficient is also encountered upon changing the enantiomer.[39, 40] As $P_1(\cos\theta) = \cos\theta$, the largest asymmetry (largest PECD effect) can be observed at $\theta = 0°$ (or 180°). This is however elusive for a non-gaseous sample because of the existence of a liquid-gas interface and the associated electron scattering inside the liquid.[12] On the other hand, the PECD asymmetry vanishes in the dipole plane (at $\theta = 90°$), which is the standard (and only) electron detection arrangement realized in currently existing LJ-PES setups. The extension to off-dipole plane detection (Figure 1a) was hence the main motivation for constructing a new setup.

## II. EXPERIMENTAL

### A. *EASI* – General features

*EASI* is a state-of-the-art setup for angle-resolved photoelectron spectroscopy from a liquid microjet, typically used in conjunction with monochromatic linearly or circularly polarized EUV to soft-X-ray radiation. For laboratory experiments, also (essentially) unpolarized radiation from a monochromatized helium plasma-discharge source, yielding the He-I α (21.218 eV), He-I β (23.087 eV), He-I γ (23.743 eV), He-II α (40.814 eV), He-II β (48.372 eV), or He-II γ (52.017 eV) emission lines, can be used. Figure 2a presents the principal arrangement of the *EASI* instrument for the case of electron detection in the floor plane – which is one of the geometries suited for $b_2$ PAD measurements – and using variable linearly polarized light (LPL). This is the standard configuration for laboratory experiments with He-I/II radiation and the most compact form adopted when moving *EASI* between the home laboratory and synchrotron-radiation facilities. In Figure 2b, a rendered graphic of *EASI* in its unique position for LJ-PECD experiments with CPL is shown. Here, the HEA (detection axis, green arrow) is tilted away from the propagation direction of the CPL (red arrow; compare Figure 1a), forming a near-magic angle of $\theta = 50°$. At this angle, the PECD asymmetry, $\sim(I_1(\theta) - I_{-1}(\theta))$, will be reduced by a factor of $\cos(50°) \cong 0.64$ from its maximum value at $\theta = 0°$ or $180°$. Due to spatial constraints, especially the dimension (size) of the HEA unit and the extension of synchrotron radiation beamline components, it was technically impossible to implement the analyzer at a smaller $\theta$-angle. Note that positioning the HEA to detect PECD electrons in the forward direction (*i.e.*, at $\theta = 130°$) is not an option because electrons cannot be detected from the far side of the liquid-jet target. This is due to the combination of strong light absorption in the dense liquid and the small electron escape depth,[1, 12] the latter making PES distinctively surface sensitive.

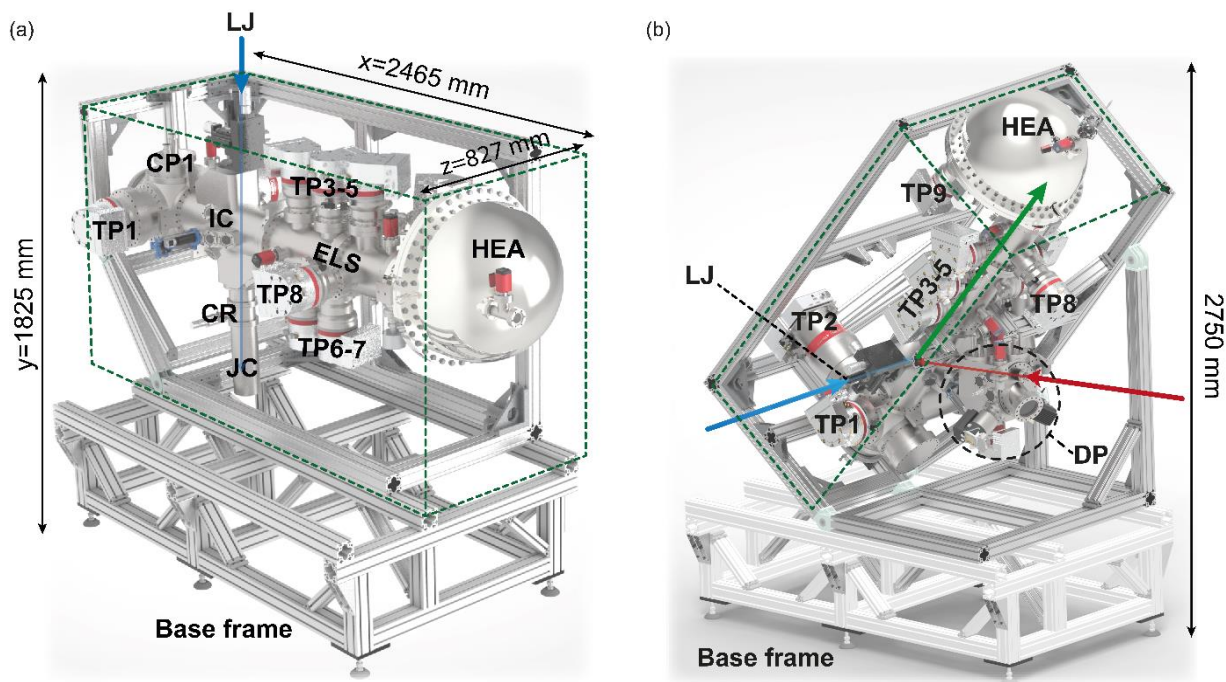

**Figure 2:** Rendered drawing of *EASI* in its most compact (smallest enclosed volume, travel) arrangement (a) and in its 'PECD-arrangement' with $\theta = 50°$ (b). In the orientation shown in (a), the liquid jet (blue arrow) travels from top to bottom. The jet direction is parallel to the entrance slit into the hemisphere. In (b), the jet enters horizontally; the HEA unit is now rotated 90° about the lens axis (green arrow) such that the entrance slit is again parallel to the jet. The most important components are labeled as follows: Interaction Chamber (IC); Electron-Lens System (ELS); Hemispherical Electron Analyzer (HEA); Turbomolecular Pump (TP#); Cryo Pump (CP); Ice Crusher (CR); Liquid Jet (LJ); Jet Catching Unit (JC). The Differential Pumping Stage (DP) will be shown in more detail in Figure 4, and the mounted VUV He-discharge light source can be seen in Figure 6. The total weight of *EASI* is 1232 kg; the weight of the base frame is 232 kg. Transformation between the *EASI* default configurations is facilitated by the compact cuboidal frame (indicated by green dashed lines), containing the core of *EASI*, which can be detached from its base frame to be freely moved in space. For each setting, a different side (face) of the cuboid sits on the lower base frame. The interaction region, *i.e.*, point of ionization, is at the same vertical distance from the floor in any orientation. Lifting, tilting, and rotating the cuboid unit is typically crane-assisted.

An important and favorable side effect of the 50°-angle being so close to the magic angle ($\theta = 54.7°$; $P_2(\cos 54.7°) = 0$) is that the angular dependence of the electron intensity on the dipolar parameter $\beta$ is suppressed (Eq. 2). This is indeed crucial when using CPL in this 'PECD' setup for achiral targets, allowing for a meaningful comparison of signal intensities arising from ionization of orbitals of different symmetry. A common application is to quantify relative intensities of different core-level peaks, often in the context of mapping solvent and solute species' distributions in solution.[1,2]

The aforementioned two other *EASI* configurations, corresponding to $\varphi = 54.7°$ (magic angle) and 90° detection within the dipole plane (compare Figure 1b), *i.e.*, in combination with horizontally LPL, are sketched in Figure 3. As explained above, measuring at just the magic angle is advantageous for many routine studies while for some measurements it is desirable to maximize or minimize relative signal contribution from a particular orbital symmetry, and this is best realized by choosing either $\varphi = 0°$ or 90°.

The modular concept of *EASI* allows for a fairly easy transformation between the various geometries. In each configuration, suitable ports allow the photon beam as well as the liquid jet to enter the interaction chamber such that the ionization spot is at the same height from the floor, not requiring any height re-adjustment at a given beamline. However, when changing between configurations the system must be vented, and several components must be re-arranged or rotated. Typically, for a given experimental run period of several days or longer, a single setting is used. Switching to another setting can be completed within 3-4 hours with the help of a crane, and the experiment can be resumed. Our experiments do not require any bake-out. Moreover, while the main experimental chamber is vented, the pressure inside the HEA and in the section containing the electron lens system is allowed to increase into the low mbar range. The necessary high-vacuum conditions for a LJ-PES experiment can be re-established within approximately five minutes.

We now consider Figure 2a in more detail, identifying the main components of the setup. These are (i) the interaction vacuum chamber (IC), which houses the liquid microjet (LJ) and is equipped with multiple cryo- (CP) and turbomolecular pumps (TP#) (their number, #, varies upon experimental demand); (ii) the electron detector, consisting of a differentially pumped electron pre-lens system (ELS) for near-ambient-pressure operation and the hemispherical electron analyzer (HEA); (iii) a multistage differential pumping unit (DP) that is only used for studies at synchrotron facilities; and (iv) a helium-discharge, high-intensity VUV source (HL), only mounted for laboratory studies. A description of the most-relevant components will be provided in the following sub-sections.

**B. Interaction chamber**

The custom-made 211-mm-diameter and 600-mm long cylindrical IC is made of grade 304 corrosion-resistant steel. A total of eleven ConFlat (CF) ports of size CF40 and four ports of size CF100 are arranged on the outer surface of the IC such that they point towards the interaction center of the chamber. Different ports can be used for photons to enter, or to mount a cylindrical or flat-surface liquid microjet and the respective jet-catcher unit depending on the specific *EASI* geometric arrangement. The intended occupancy of the ports in a given geometry is indicated in Figures 2 and 3 by the red (photon-beam axis), green (electron-detection axis), and blue (LJ-flow axis) arrows/dots.

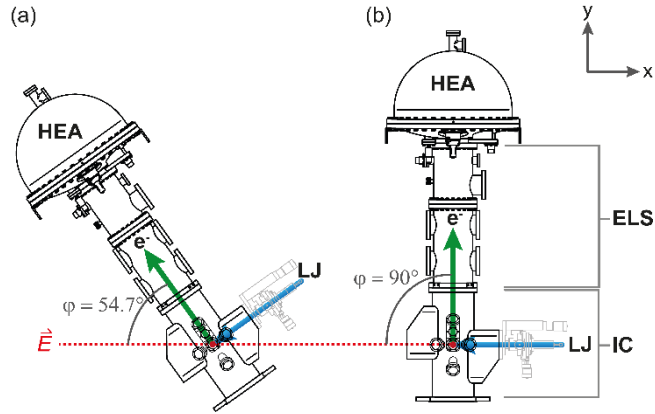

**Figure 3:** Sketches showing *EASI* without the cuboidal frame in the $\varphi = 54.7°$ (a) and $\varphi = 90°$ (b) configurations in the x-y (dipolar) plane defined in Figure 1b. As in Figure 2, in each configuration, electrons are detected perpendicular to the flow direction of the liquid jet. Red dots indicate the ionization point associated with synchrotron radiation propagating perpendicular to the figure plane, away from the observer. The horizontal electric field vector $\vec{E}$ (in the case of horizontally LPL) is also indicated.

In order to effectively shield induced and the Earth's magnetic field at the interaction point (red dots in Figure 3), which is of major concern for the quantitative detection of low-kinetic-energy electrons, we mounted an additional µ-metal superalloy shielding (µSH), a cylindrical inset within the cylindrical IC that forms an inner layer over its entire length. Typical magnetic fields measured at the interaction point of liquid jet and light are approximately 0.5 µT for the horizontal and 0.3 µT for the vertical components. The µSH incorporates thirteen 30-mm and two 40-mm diameter holes on its surface, positioned such that the ports on the IC have an unobstructed view on the interaction point. The two larger ports are used for the liquid jet and provide enough space for its positioning. Fixations of the µSH shield are made of titanium. At one end, the IC connects via a CF200 flange to the magnetic shield of the lens system of the analyzer, giving *EASI* the elongated appearance. A view into the IC and the µSH, along the cylinder and electron detection axis of the hemispherical energy analyzer is shown in Figure 6a.

In order to achieve a sufficient vacuum base pressure in the IC, approximately $5 \cdot 10^{-5}$ mbar for typical liquid-jet experiments, two turbomolecular pumps (TP1, TP2; see Figure 2a) are mounted approximately 400 mm downstream of the analyzer orifice. TP1 is a 1360 l/s (Pfeiffer ATH 1603 M) TP, and TP2 is a 790 l/s (Pfeiffer HiPace® 800 M) TP. Each TP is backed by one corrosion-resistive 10 l/s scroll pump (Edwards xDS35i C). Main pumping of the evaporating liquid jet is however accomplished by at least one additional cryo pump (CP). When operating a cylindrical microjet in the laboratory, we use a single CP, consisting of a cylindrical liquid-nitrogen (LN$_2$) trap made of stainless steel with a surface area of ~1000 cm$^2$. This is CP1 in Figure 2a. The pumping speed of CP1 for water vapor is approximately 10,000 l/s for a pristine trap surface,[41] *i.e.*, exceeding the capacity of TP1 and TP2 by nearly an order of magnitude. If required, for instance when operating a liquid flatjet with much higher evaporation rates, and for synchrotron-light experiments, up to three LN$_2$-traps can be added. Alternatively, a recirculating system for liquid collection can be used instead of a liquid nitrogen cold-trap, as we detail in Section II. E.

**C. Electron detection**

The electron analyzer used with *EASI*, a Scienta Omicron HiPP-3, is a high-energy-resolution state-of-the-art HEA with 200 mm central radius of the hemisphere. It has rather similar properties as its predecessor, which has been described in detail previously.[42] Here, we review the main features, and highlight several new ones. One characteristic is the separate pre-lens that provides efficient differential pumping, in conjunction with two further differential-pressure stages within the HiPP-3 lens system, and the electron optics for PES imaging. The HEA can operate over a large pressure range, including typical $10^{-4}$-$10^{-5}$ mbar pressures under standard liquid-jet conditions, but also sustaining pressures as high as 5-30 mbar in the IC. This enables the probing of liquid surfaces other than those associated with liquid jets or jets within some gaseous (near-ambient-pressure) environment. To provide the required vacuum conditions, each stage of the electron lens is pumped by two 255 l/s turbomolecular pumps (Pfeiffer HiPace® 300 M), labeled TP3, TP4 etc. in Figure 2a, with each given pair of TPs being backed by one 10 l/s scroll pump (Edwards xDS35i C).

The pre-lens is equipped with a small front aperture (orifice or skimmer, representing the first differential pumping stage) at the tip of a graphite-covered titanium analyzer cone, that permits electrons to enter the analyzer. The opening angle of the front cone is 45° (see Figure 6b below), which is sufficiently steep to position the liquid jet in close proximity. The slim front-cone design also provides sufficient space for the exit capillary of the He-discharge source, requiring a short working distance (see below).

Different orifice sizes are available, although we almost exclusively use an 800-μm orifice for liquid-jet experiments. This small opening allows for an elevated maximum pressure in the IC and at the same time effectively protects the lens system from contamination arising from the volatile-sample environment. The acceptance angle is approximately ±15°, with the accurate value depending on the retardation ratio, eKE/$E_p$ (see below). In all cases, this angle is smaller than the ±26.6° geometric acceptance for the nearly point-sized liquid-jet sample in front of the 800-μm orifice. The HiPP-3 analyzer is capable of covering a ~2-1500 eV electron kinetic-energy detection range. Extension to even higher kinetic energies can be achieved when using a higher-voltage power supply. The (pre-)lens design – having a first skimmer followed by a second one – enables operation of the analyzer in two different modes, the swift acceleration mode and the normal transmission mode. For realization of the former, the second skimmer is held at a potential, while it is grounded for the normal mode. In the swift mode, electrons are thus accelerated as soon as they enter the analyzer, which greatly reduces the inelastic scattering of the photoelectrons with the dense water-gas environment, thus enhancing the transmission at near-ambient-pressure conditions. In addition, this mode increases the angular acceptance (an aspect less relevant when using a small entrance-cone orifice) in both ultra-high vacuum and mbar pressure conditions. In combination, this leads to an increase in signal of up to a factor of ten compared to traditional lens modes.

The energy resolution as well as the electron transmission are determined by the size of the entrance slit into the hemisphere (selectable between 0.1-4.0 mm using nine different straight slits) and by the pass energy, $E_p$, with the latter being restricted to pre-set values depending on the lens mode used. For instance, in the transmission mode, $E_p$ can be selected from 20, 50, 100, 200, and 500 eV, covering an electron kinetic energy range of 20-1500 eV. With 20 eV pass energy and 500 eV kinetic energy an energy resolution better than 15 meV full width at half maximum (FWHM) is specified; note however that in the case of aqueous solutions, most PES peaks are considerably wider due to an intrinsic distribution of hydration / solvation configurations of different energies. Other available modes include the angular (±9° parallel angular range; 100-1500 eV kinetic-energy range) and spatial (20-1500 eV kinetic-energy range) modes. The latter mode is specified to achieve a spatial resolution <10 μm for kinetic energies below 1200 eV. For measurements from a cylindrical liquid microjet the spatial mode is of little relevance, but this mode will be exploited in upcoming characterizations of planar microjets where several properties (including jet thickness, solute concentration, and temperature) might vary when probing the liquid sheet along the flow direction.[43, 44]

Another unique lens mode of the HiPP-3 is the UPS upgrade, which enables low-kinetic-energy measurements with high energy resolution and a simultaneous increase of electron transmission. This adds the following features: Energy resolution <5 meV FWHM at 5 eV pass energy and 10 eV kinetic energy; pass energies 2, 5, 10, and 20 eV; kinetic-energy ranges of near-zero to 60 eV (UPS mode), near-zero to 100 eV (angular mode), near-zero to 20 eV (spatial mode). It is the former mode, typically using $E_p$ = 20 eV, which is indispensable for the near-threshold measurements, *i.e.*, the main mission of *EASI*. Here, the detection of electron energies smaller than $E_p$ is accomplished by a custom-made lens table developed by Scienta Omicron.

The actual electron detector unit at the exit of the hemisphere consists of a stack of two 40 mm-diameter microchannel plates (MCPs) in a Chevron arrangement, combined with a phosphor screen (type P46) to image the position of electron hits in two dimensions. The screen image is recorded through a viewport by a CCD camera (Basler scA 1400-17gm; acquiring 17 frames per second) on the non-vacuum side of the hemisphere. A rectangular section of this image is divided into a maximum of 1064 energy channels in the energy-dispersive and 1000 channels in the non-dispersive direction, which for some lens modes corresponds to a spatial or angular coordinate at the interaction center. The camera may either count illuminated pixels above a certain threshold to determine the number of detected electrons (pulse-counting-mode) or generate the spectrum from interpreting the gray-scale levels of the CCD image (ADC-mode). To obtain the absolute count rate from the recorded ADC spectra, a calibration factor (multiple counting factor, MCF) must be determined before measurements, individually for each pass energy. In routine operation, we use the gray-scale mode.

### D. Differential pumping chamber

In synchrotron-radiation measurements employing a liquid jet, a highly efficient differential pumping unit (DP) must be placed between the IC and the beamline since in the latter a pressure requirement of $10^{-9}$ mbar to $10^{-10}$ mbar usually must be met (as compared to the typical $10^{-4}$ mbar to $10^{-5}$ mbar pressure in the IC, or considerably worse, $10^{-3}$ mbar, in the case of flatjets or near-ambient-pressure studies). For *EASI*, we designed a novel, highly compact three-stage DP, which is shown in Figure 4. The total length is only 355 mm, allowing accommodation of this unit even at beamlines with a relatively short focal length. Each stage is pumped separately. The first one, close to the interaction chamber (low-vacuum side), is pumped by one 255 l/s TP (Pfeiffer HiPace® 300 M), while the other two stages are each pumped by one 90 l/s TP (Leybold TURBOVAC 90i). All three pumps are backed by a single 10 l/s scroll pump (Edwards xDS35i C). In order to increase the pumping speed and to more efficiently pump water vapor, we additionally use two $LN_2$ traps (cryo pumps; CP in Figure 4), each with a surface area of ~580 $cm^2$. With that we maintain a $10^{-9}$ to $10^{-10}$ mbar pressure in the connecting beamline chamber even for mbar-range pressure in the IC.

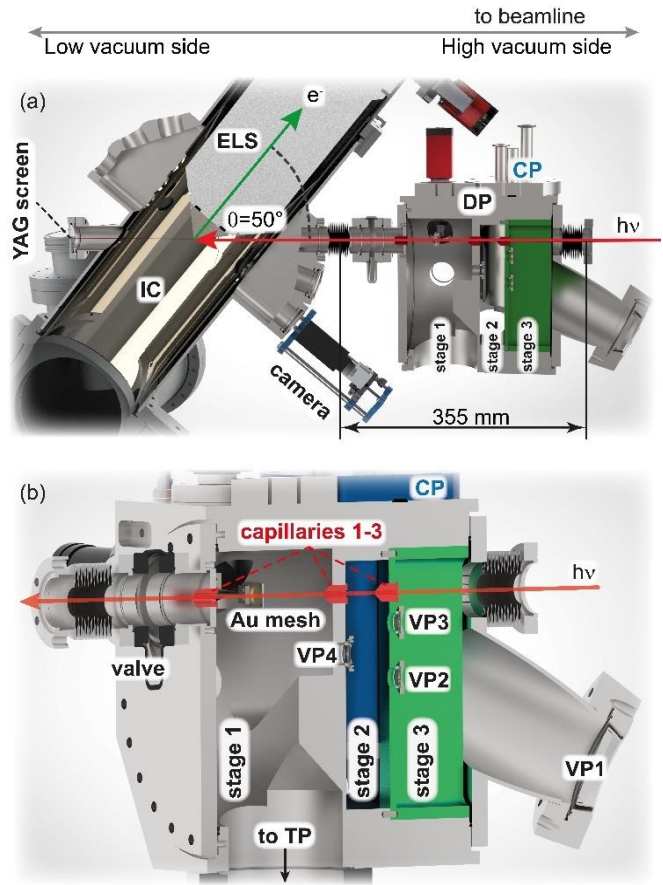

**Figure 4:** (a) Sketch of the differential pumping unit (DP), including dimensions and mounting orientation onto the IC in the 'PECD' configuration (compare Figure 2b). (b) Close-up of the DP. The main components are labeled as introduced in the text: Stages 1-3; view ports 1-4 (VP1, VP2 etc.); capillaries 1-3; camera; cryo pumps (CPs; only one can be seen from this viewing angle). A single-crystal cerium (III)-doped yttrium aluminum garnet (YAG:Ce) screen of 0.1 mm thickness and 20 mm diameter is placed on the far side of the IC for visual inspection of the beam position, as it emits visible light when hit by UV light or X-rays.[45]

The photon beam propagates through the DP via three 20-mm long stainless-steel capillaries, which connect the stages. On the high- and low-vacuum sides, we use a capillary of 3 mm and 8 mm inner diameter, respectively; the capillaries can be easily exchanged if required by the experimental conditions. To aid alignment of the whole unit, we coated the ends of the capillaries facing the beamline with fluorescence powder (Honeywell LUMINUX Green B 43-3), and the green-glowing spot allows the position of the light beam to be tracked and observed through dedicated viewports (VPs).

Two further elements for beam monitoring are mounted inside the IC: A retractable gold mesh (Precision Eforming, 333 LPI) for quantitative monitoring of the photon flux shortly downstream of the interaction point and a YAG:Ce plate for visual inspection of the photon beam shape.

## E. Liquid jets and alignment

Vacuum liquid microjets are produced by pushing water (or other solvents or solutions) through a micron-sized orifice into vacuum.[1, 4, 46] We usually use 15-30 µm inner-diameter quartz-glass capillaries of ~3 cm length, made in-house, to obtain cylindrical microjets. At times we also use platinum plates (30 µm inner diameter; 2 mm outer diameter), similar to what has been reported in our early LJ-PES studies.[47, 48] Resulting jet velocities are in the 20-80 ms$^{-1}$ range, depending on the given experimental conditions. More recently, we have also generated planar-surface microjets by colliding two cylindrical jets at a suitable angle, analogous to the design described in Ref. 44. Several different capillary materials, including quartz, have been tested. In the exemplary photoelectron spectra from a liquid-water planar jet which will be presented below, 65-µm inner-diameter polyether ether ketone (PEEK) tubes were used, at a 46° collision angle.

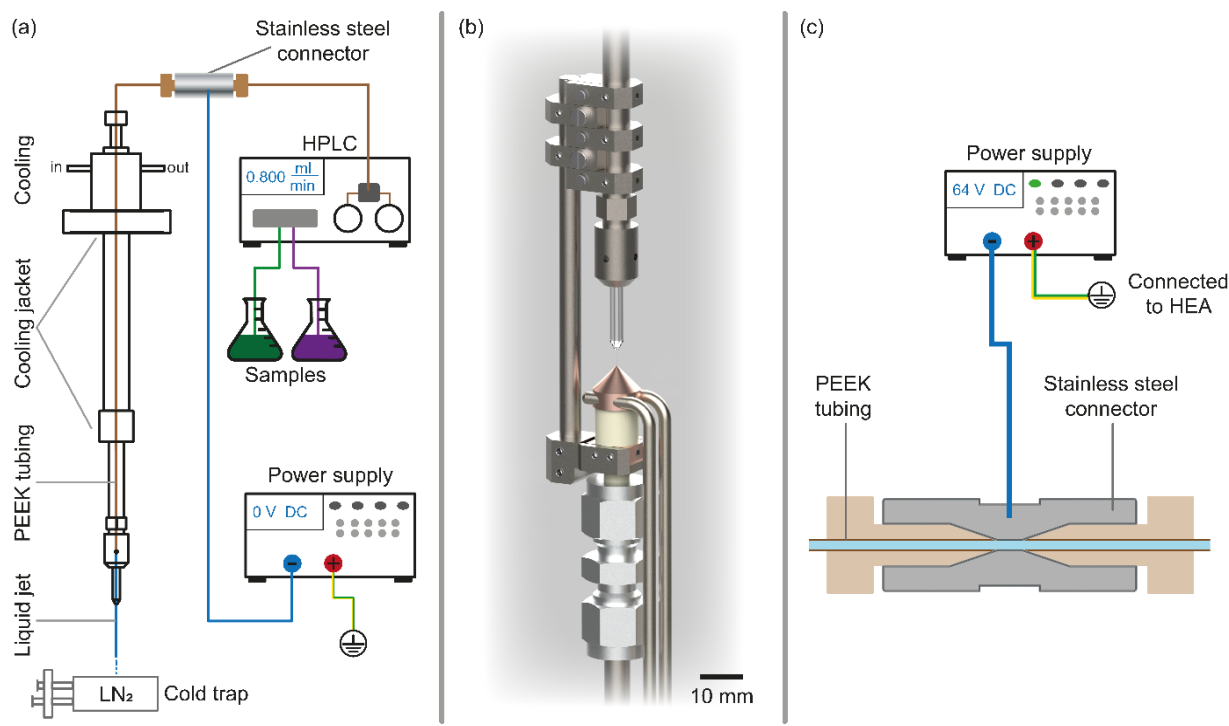

**Figure 5:** (a) Schematic of the standard liquid-jet setup, showing the jet rod, HPLC pump, liquid-nitrogen cold trap and electric connections for biasing or grounding the liquid jet. (b) Rendered graphic of a liquid-jet catcher / recirculation unit. The liquid-jet-injecting glass capillary and the jet catching cone with 500 µm orifice at <7 mm distance from the capillary tip in the direction of the flowing jet, are mounted on a common support. Mutual jet and catcher positions are mechanically adjustable. The bronze catcher cone is typically held at 80°C (associated heat pipes are shown) to prevent clogging of its orifice upon water ice formation; details of pumping on the catcher side are not shown. This slim and compact design (using several components of Microliquids design, now Advanced Microfluidic Systems GmbH – AdMiSys)[49] fits into the same ports of the cylindrical shielding (µSH) used for jet operation without a recirculating unit. (c) Schematic of the stainless-steel connector which is in contact with the aqueous solution. For grounding the liquid jet, the connector is linked to the grounded HEA, and for biasing the jet, the connector is electrically linked to a power supply.

The liquid to be pushed through the capillary is pumped through PEEK interconnected tubing of different inner diameters, 130 and 800 µm, by a high-performance liquid chromatography pump (HPLC; Shimadzu LC-20AD), equipped with four inlets to accommodate quick switching between different solutions. A sketch of our standard liquid-jet setup is shown in Figure 5a. The HPLC takes in filtered solutions channelled via an in-line Shimadzu DGU-20A$_{5R}$ degasser to enable the simultaneous preparation of different solutions. The high-pressure side of the HPLC is connected to the jet holder via approximately 5 m of PEEK tubing which is interrupted by several Teflon® inter-connecting tubing segments of different inner diameter, as experience showed this to dampen occasional oscillatory throughput variations of the HPLC operating in the low-pressure regime. Under these conditions, typical flow rates are 0.4 ml/min up to 1.5 ml/min, corresponding to 5-30 bar pressure in the tubing for (cylindrical) liquid-jet experiments at solution temperatures of 10 °C. In a given set of experiments, the flow rate is typically constant, adjusted by HPLC backing pressure, which depends on solution viscosity and temperature as well as jet diameter. Control and stabilization of the jet temperature is accomplished by flowing a water-ethanol volume mixture (30:70) through a cooling jacket of the jet rod. Towards this end, a closed-flow temperature-stabilized cycle is maintained by a chiller unit (Julabo CORIO CD-200F). Typically, the temperature is set between 4-20 ºC, depending on the experiment. A small, but unquantified difference of the set temperature to the actual temperature at the point of expansion may occur because the cooling jacket ends few centimetres before the actual nozzle.

In vacuum, the produced laminar liquid microjet quickly cools by evaporation. Eventually it disintegrates into droplets and freezes, and the resulting spray is collected downstream of the flow. For a jet traveling horizontally, we typically use a regular LN$_2$ cold trap of similar design as the cryo pumps described above.[3, 47] In case the jet travels vertically from top to bottom (see Figure 2a), it is terminated by a steel cylinder submerged in a liquid-nitrogen bath. A noteworthy technical detail is that at some suitable position, approximately 100 mm upstream of the respective catching unit, a motorized rotating wireframe, with a shape resembling a kitchen mixer, is placed (CR in Figure 2a). This unit nebulizes the liquid flow and prevents ice needles growing back from the cold-trap surface towards the jet capillary, *i.e.*, opposite to the flow direction. This also helps to maintain a rather homogenous coverage of the cold surface which slows down the decrease of pump efficiency and considerably extends the available measurement time between venting and cleaning cycles.

The liquid-jet rod, the supporting metallic unit for a cylindrical single jet, consists of an inner tube with a socket to hold the quartz capillary, PEEK tubing, and an upstream connector (see Figure 5a). These parts are sleeved with an outer tube to stabilize the construction, also acting as a jacket for the coolant liquid. This whole unit, a modified Microliquids design (now Advanced Microfluidic Systems GmbH – AdMiSys),[49] is mounted on a high-precision x-y-z-manipulator (Hositrad, MA2000 series). All parts of the jet-assembly unit that immerse into the magnetically shielded region of the IC are made of titanium or other non-magnetic materials, typically tungsten, copper, or aluminium. Parts (except for the quartz capillary) in the vicinity of the ionization region have been graphite-coated to assure a common electric potential. All of these parts are fully electrically insulated from the liquid sample solutions. As mentioned above we can alternatively operate the liquid jet as a recirculating system, based on collecting the liquid jet before freezing. Our system is similar to those previously reported,[50, 51] consisting of a catcher, of approximately 1 cm$^3$ size, made of bronze which is connected via stainless-steel tubing to a solution reservoir container; see schematic in Figure 5b. The liquid jet shoots into the 500-µm orifice of the cone of the catcher after <7 mm travel in vacuum. The advantage of a recirculating unit, other than recycling or recovering the solution, is the reduction of pressure in the main IC and, even more important for our experiments, the deposition of volatile species from the solution on the chamber walls can be reduced to achieve more temporally stable vacuum conditions.

Given the micron-sized diameter of the liquid jet, highly accurate positioning of the jet is mandatory, and is accomplished using a high-precision x-y-z manipulator (Hositrad, MA2000 series), modified by the manufacturer to achieve a spatial resolution of 2.5 µm with a repeatability of 1.25 µm. To visually monitor the liquid-microjet performance and its position, we use two Basler acA2440 – 35-µm cameras in combination with suitable telescopes. One camera (equipped with a NAVITAR NMV-100 objective and a 15-mm spacer) is aligned to the center-symmetry axis of the electron-analyzer lens, and observes whether the liquid jet is centered in front of the HEA first skimmer. The typical jet-to-skimmer distance is 500-800 µm to match the imaging distance of the HEA lens system; at the same time this distance corresponds to a suitable electron transfer length between liquid jet and analyzer at the typical $10^{-4}$ to $10^{-5}$ mbar water vapor pressure in the IC.[1] A view seen by this camera, although in the presence of a flatjet sample, is shown in Figure 6a. The other camera (combined with a RICOH FL-BC7528-9M objective) is directed at the jet and the HEA entrance cone at an angle perpendicular to the detection axis (see Figure 4). With this combination of two cameras, we can accurately re-position the jet for each measurement, in any of *EASI*'s geometric arrangements.

In order to obtain meaningful liquid-jet photoelectron spectra, we have to assure that the solutions have a sufficient electrical conductivity. Neat liquid water and many aqueous solutions containing no ions are however poorly conductive, and a tiny amount of salt needs to be added, as discussed previously.[1] Another effect of salt addition is the compensation of electrokinetic charging of the jet surface.[52, 53] In a related context, molecular dipoles at the solution surface can also give rise to surface charging. Any quantitative information on the energetics (absolute binding energies) of ionized solute and solvent then requires that the jet is either properly grounded to the HEA, or that a stable bias voltage is applied, as recently discussed.[12, 23] To connect the liquid jet to the electrostatic potential of the analyzer, we have inserted a stainless-steel through-connector (as used with HPLCs) in the high-pressure side of the PEEK line, at a few tens of millimeters upstream of the liquid-jet holder. This design, detailed in Figure 5c, turned out to provide a much lower contact resistance compared to earlier versions, in which the electrical contact to the solution was provided by a gold wire. For PES experiments from a biased liquid jet, the solution is fully electrically insulated from any other potential, and only connected to a high-precision power supply. We use a Rohde & Schwarz HMP 4030 high-precision voltage source, or for higher voltages (60-300 V) a Delta Electronics ES 0300-0.45 power supply.

**F. Helium lamp**

A helium plasma-discharge source (Scienta Omicron VUV5k) enables LJ-PES valence measurements in the laboratory. Here, we greatly benefit from the aforementioned HEA VUV lens-mode. This combination has been recently applied to determine absolute lowest ionization energies of water and solutes.[23, 54] The VUV5k, equipped with differential pumping, is operated with helium 6.0, and liquid nitrogen-cooling of the connecting gas-line removes water residuals and contaminant gases. A given discharge line – we primarily use He-I α (21.218 eV), He-II α (40.814 eV), and He-II β (48.372 eV) – is selected by an 80 × 30 mm$^2$ toroidal grating with 1200 lines/mm. The monochromatic radiation is then directed into a collimating 300 μm-inner-diameter (75-mm-long) glass capillary, producing a 300 x 300 μm$^2$ focus at 5 mm focal length, which corresponds to the distance between capillary exit and liquid jet (see Figure 6). The total photon flux at the ionization region, without grating and focusing capillary implemented, is approximately $3 \cdot 10^{14}$ photons/s, and the flux of the focused He-I α photon beam, using the 300-mm capillary, is specified as ~1.8 $10^{11}$ photons/s. The He-II α photon flux is ~$1 \cdot 10^{10}$ photons/s. Both values are sufficient to acquire high-quality LJ-PE spectra;[12] an example of which will be presented below. The He-light source is dismounted when performing experiments with synchrotron radiation.

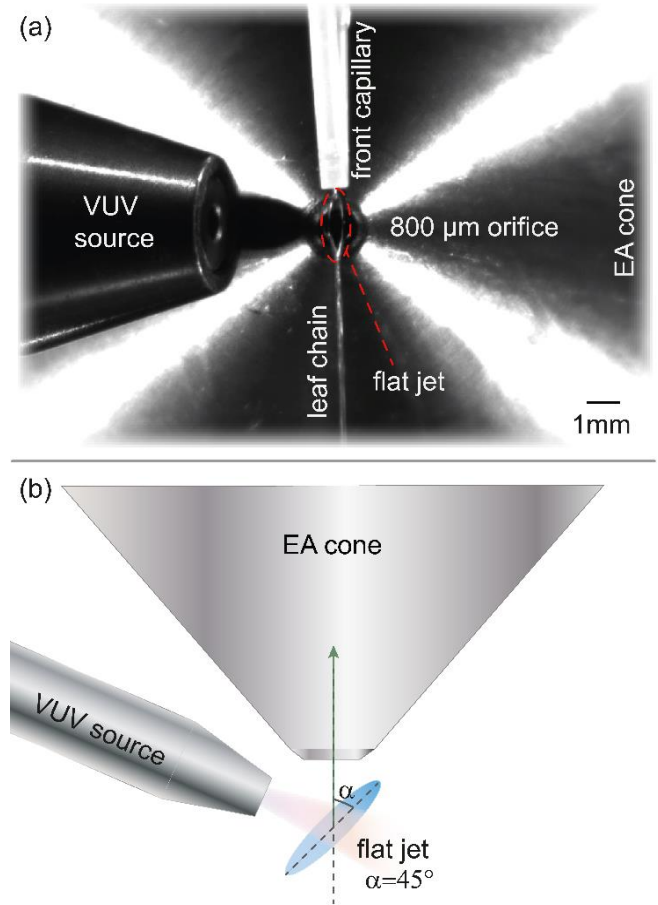

**Figure 6:** (a) Photograph of the view into the IC along the lens axis of the HEA, and centered on the analyzer cone with its 800-μm orifice. In front of the orifice, a water flatjet, 1.2 mm long and 0.6 mm wide, is seen. The thickness of the jet is approximately 20-25 μm. A subsequent chain of pairwise orthogonal leaves, forming downstream along the flow direction, is not resolved by the camera used here for jet alignment. The flatjet is formed by two colliding cylindrical jets, each with 65 μm inner diameter and a 46° collision angle; simultaneously changing the diameters of the cylindrical jets can be used to adjust size and thickness of the planar jet. Above the flatjet, one of the two PEEK capillaries generating the colliding cylindrical jets can be seen. At the left-hand side, the focusing capillary, receiving light from the VUV discharge source, is shown at a working distance of approximately 5 mm. The angle between the detector axis and the VUV-photon beam is 70°. (b) Schematic of a top view of the same situation, showing the angle α between the surface plane of the first flatjet leaf and the direction of electron detection.

## III. *EASI* PERFORMANCE

This section presents selected aspects of the *EASI* instrument performance, exemplified for two applications that qualitatively extend the liquid-phase photoemission studies that had been reported hitherto. We first underpin the feasibility of studying liquid-phase PECD using *EASI*. In the second example, we present the outer-valence spectra of neat liquid water and of a sodium-iodide aqueous solution, here recorded from a planar liquid microjet.

## A. PECD in the C 1s spectrum of gaseous and liquid fenchone

Fenchone is a chiral organic molecule, a terpene, consisting of a six-membered carbon ring stabilized by an additional carbon bridge, which contains two stereo centers, as depicted in the inset of Figure 7. A single ring site is double-bonded to an oxygen (carbonyl group). The chiral centers are located at the positions connecting the ring and bridge; these sites are labeled with a * in Figure 7. This rigid structure of the molecule makes fenchone a suitable, prototypical system for studies of PECD, as isomerism plays a minor role; isomerism might otherwise considerably complicate the interpretation of observed emission asymmetries. After the discovery of a significant gas-phase PECD in its C 1s spectrum,[18] fenchone has been used in a number of valence-level PECD studies[55, 56] as well as for exploration of more sophisticated dichroic effects upon multi-photon excitation.[57-61]

Another noticeable aspect in the present context is that fenchone is a liquid at room temperature, with somewhat higher viscosity than liquid water, and approximately 10-times lower vapor pressure at room temperature. Although the formation of a liquid-fenchone jet is almost as straightforward as for water, the immiscibility of fenchone with water requires that all surfaces of the jet setup in contact with the liquid must be completely water-free to avoid clogging of the quartz capillary. Unlike water, fenchone is a nonpolar molecule, which has important consequences for the interpretation of liquid-jet spectra, as discussed below.

Commercial samples of (1R,4S)-(−)-fenchone (Sigma-Aldrich, ≥98% purity) and (1S,4R)-(+)-fenchone (Sigma-Aldrich, ≥ 98% purity) were used in our experiments. To avoid charging of the jet (compare Section II. E), the conductivity of liquid fenchone was increased by adding tetrabutyl-ammonium nitrate salt (TBAN) to 75 mM concentration to the liquid. We used liquid flow rates between 0.7 and 1.0 ml/min through a 28 μm glass capillary; the bath temperature upstream of the nozzle was set to 10 °C. Very similar conditions are applied in typical LJ-PES studies from water and aqueous solutions. The liquid jet was kept at ground potential, unless otherwise stated. Experiments were conducted at the soft-X-ray beamline P04 of the PETRA III storage ring.[62] Opposite-helicity C 1s PE spectra were sequentially recorded, repeatedly switching between *l*- and *r*-CPL; here, we follow the 'optical' convention helicity $p = +1, -1$.[31] A fairly large X-ray focal spot size, 180 μm horizontal (parallel to the liquid jet) and 30-40 μm vertical, was deliberately chosen in order to minimize electron signal sweep-to-sweep fluctuations that may arise from small liquid-jet instabilities. We collected between 10 and 30 spectral sweeps for each helicity of the radiation, with the acquisition time of a single sweep being between 30-60 s. The HEA was operated in the novel VUV lens mode at a pass energy of 20 eV. As the absolute helicity of the undulator radiation had not been previously determined, this information could be inferred in the present study. We identified that a negative shift of the opposing magnetic arrays of the APPLE-II undulator at the P04 beamline corresponds to *l*-CPL by comparing measured asymmetries (see Figure 7) to previous gas-phase fenchone results.[18]

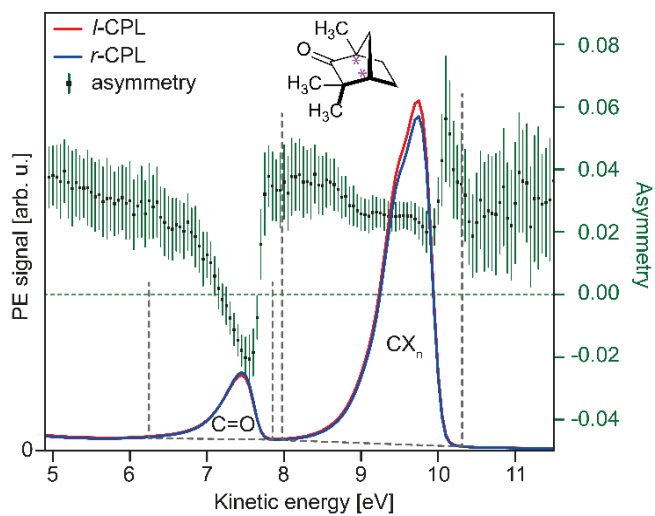

**Figure 7:** Photoelectron spectra of gaseous 1R-fenchone recorded at a photon energy of 300 eV with *left-* and *right-*handed circularly polarized light. The asymmetry, $A$, of the spectra derived for every data point (symbols) is not symmetric about zero because the intensities of the two different photon beams are not identical, see text for details. Two chemically shifted C 1s components near 7.4 eV (C=O site) and 9.7 eV kinetic energy (all $-CX_n$ sites; X includes carbon and hydrogen atoms) with different asymmetry are seen. The chiral centers of fenchone are labeled with * in the molecular sketch. Vertical and nearly horizontal lines indicate area intervals used for peak area determination and the respective background, see text for details. Error bars of the asymmetry designate the standard deviation of the mean values, derived from the variation of the individual sweeps. The modulation of $A$ near 10 eV kinetic energy, exhibiting a small minimum towards smaller $A$ values, with a sharp overshoot at slightly higher kinetic energy arguably results from minute differences of the peak profiles, reflecting the slight disturbances from fluctuating potentials of a charged liquid jet in the present experiment.

**Gas-phase fenchone C 1s spectra** – C 1s PE spectra from gas-phase 1R-fenchone, collected from the surroundings of the liquid jet at 300 eV photon energy are presented in Figure 7. Signal intensities in the figure are displayed as measured. Photon energies in the present work were calibrated by a standard procedure which fixes the grating pitch angle for specular reflection, and are estimated to have ±0.2 eV accuracy in the energy range studied. Measured kinetic energies were corrected to match the binding energies (BE) of the two main peaks, previously reported in the literature.[18] Note that the photon energy of 300 eV corresponds to approximately 7.4 eV kinetic energy of the photoelectron due to ionization of the carbonyl carbon (~292.6 eV BE) at which the largest PECD effect has been reported.[18] Two spectrally well-separated peaks are observed in Figure 7, in very good agreement with earlier results.[18] The more intense peak near 9.7 eV KE (~290.3 eV BE) and denoted as $-CX_n$ arises from the nine energetically overlapping carbons featuring C–C or C–H bonds. The small peak is due to C 1s ionization of the carbonyl group (C=O). An important implication of the gas-phase work on core-level PECD is that the emergence of a measurable asymmetry generally is not restricted to atoms forming the chiral center of a molecule.[35] A PECD effect for the C=O group may therefore be expected, and is discussed in detail in Ref. 18.

The intensities of spectra obtained with *l*-CPL and *r*-CPL are displayed as measured in Figure 7. For a meaningful discussion of their intensity differences, it is useful to discuss them in a normalized form. We therefore display the channel-by-channel asymmetry *A*, defined as ($I$(*l*-CPL)–$I$(*r*-CPL))/($I$(*l*-CPL)+$I$(*r*-CPL)). This quantity shows a fairly constant behavior over the –$CX_n$ peaks and a strong trend towards the opposite sign of *A* in the region of the non-chiral C=O carbon (Figure 7). In order to further interpret these results, we make use of the finding of Ulrich *et al.*[18] that the combined intrinsic asymmetry of the –$CX_n$ peaks cancels out to a good approximation. Indeed, within the error bars, it is hardly different from the background asymmetry. We can then use our measured value of *A*, which corresponds to the observed overall shift of *A* by about ~0.025 towards positive values, to quantify the instrumental asymmetry which is caused by a slight imbalance between intensities of the APPLE-II undulator at positive and negative shift. In Ulrich *et al.*,[18] a sophisticated scheme involving rapid alternation between two beams of opposite helicity, produced in a twin undulator, was used to achieve cancellation of these apparatus effects.[24] However, the asymmetry of negative sign for the carbonyl C 1s does arise from PECD. Below, we will discuss its values, corrected such that the apparatus asymmetry of the –$CX_n$ peaks vanishes. A similar procedure was used in an early work on gas-phase core-level PECD.[35] The low-kinetic-energy flanks of both peaks are caused by unresolved vibrational structure.[63] The slight increase of the value of *A* along both flanks towards lower kinetic energies is caused by a small, constant background present in both spectra but with slightly different intensity, the influence of which increases with decreasing signal intensity.

In Table 1 we compare results for the intrinsic chiral asymmetry parameter $b_1^{+1} = A/\cos(130°)$ extracted from the data shown in Figure 7, and additional measurements at photon energies of 301 eV, 302 eV, and 305 eV after correction for the instrumental asymmetry, to results in the literature. Measurements from both enantiomers were directly compared at 300 eV in the gaseous phase, and at 302 eV for the liquid (yet to be discussed in detail). Values of $b_1^{+1}$ in the present work were determined from peak areas minus a linear background between the respective vertical lines shown in Figure 7.

**Table 1:** PECD parameter $b_1^{+1}$ of the carbonyl C 1s photoemission line of both enantiomers of fenchone measured at different photon energies. Results for the gaseous phase are compared to the literature.[18] Literature values in parenthesis were obtained by interpolation between the nearest reported energies. An uncertainty of approximately 0.004-0.007 in the values for $b_1^{+1}$ is found from the sweep-to-sweep fluctuations in our study and corresponds to the approximate size of error bars shown in Ref. 18. 1R-Fenchone is used as a shorthand notation for (1R, 4S)-(-)-

fenchone, and 1S-Fenchone for (1S, 4R)-(+)-fenchone. The value given for liquid phase fenchone has not been corrected for the presence of gaseous components, see the main body of the text for details.

| Phase | Enantiomer | KE [eV] | $h\nu$ [eV] | $b_1^{+1}$ (this work) | $b_1^{+1}$ (lit.)[18] |
|---|---|---|---|---|---|
| gaseous | 1R-Fenchone | 7.4 | 300 | 0.067(6) | 0.072, 0.078 |
| | 1S-Fenchone | 7.4 | 300 | -0.084(5) | -0.069, -0.086 |
| | 1S-Fenchone | 8.4 | 301 | -0.085(7) | (-0.075) |
| | 1S-Fenchone | 9.4 | 302 | -0.050(4) | (-0.055) |
| | 1R-Fenchone | 12.4 | 305 | 0.019(4) | 0.025 |
| liquid + gas-phase | 1R-Fenchone | 9.4 | 302 | 0.02 | - |
| | 1S-Fenchone | 9.4 | 302 | -0.02 | - |

The results of our few-photon-energy measurements and of the respective literature values, presented in Table 1, are found to be in reasonable agreement. Note that both experiments reveal the change in sign of $b_1^{+1}$ for 7.4 eV kinetic energy (300 eV photon energy) when switching between the two fenchone enantiomers. We also point out that the fenchone –$CX_n$ peak is somewhat better resolved in the present study, exhibiting a more structured peak top. Combined with much faster data acquisition, this opens the perspective to study sub-structures within the –$CX_n$ peak of fenchone, or of congested line systems of other chiral molecules.

**Liquid fenchone C 1s spectra** – Turning now to analogous measurements from liquid fenchone, we expect that quantification of PECD will be complicated by (photo)electron scattering which occurs in any condensed matter system. Inelastic and quasi-inelastic scattering leads to the formation of an intense background at low kinetic energies,[12] which inevitably overlaps with photoelectron peaks when measured near a given photoionization threshold energy (as is the case here). We have recently shown for neat liquid water that the presence of such scattering channels significantly perturbs the native photoelectron peak shape and also affects the peak center position, making a meaningful determination of the energetics futile below approximately 10 eV kinetic energy.[12] An exact value of this limiting kinetic energy for liquid fenchone has not been determined but indeed, the existence of a strong scattering background is a major challenge for PECD measurements from liquid fenchone and aqueous solution alike.

C 1s PE spectra of a grounded liquid jet from 1R-fenchone, measured at 302 eV photon energy with *l*- and *r*-CPL, respectively, are shown in Figure 8. At this photon energy, the lower-kinetic-energy C 1s peak of C=O from liquid fenchone can still be resolved at ~9.5 eV KE atop of the inelastic background. At a slightly lower photon energy of 300 eV, as used for the measurements on gas-phase fenchone (Figure 7), the C 1s peak of C=O could not well be separated from the large scattering background (not shown here), illustrating the detrimental impact of electron scattering at these low energies. Noticeably, the liquid-phase spectra exhibit a similar effect of the CPL helicity on the intensity. Again, the main C 1s peak exhibits higher intensity for *l*-CPL originating from the same instrumental effect, but the respective flipped intensities of the C=O peak are only revealed upon spectral analysis. Performing the analogous analysis as for the gas phase, we find the respective liquid-phase asymmetry values which are presented in Table 1. These values have the same sign as the gas-phase asymmetries but are smaller by a factor of approximately 2.5. Quantitative results strongly depend on the exact choice of the background model though, with one possible approach sketched in Figure 8. A more elaborate analysis will be presented in a forthcoming work. We also note that our simple analysis does not account for possible ordering of molecules at the liquid–vacuum interface, in which case the angular distributions may be affected considerably.[38] Such a contribution cannot be quantified here but a previous PAD study from liquid water[64] suggests that for a cylindrical liquid jet, orientation effects average out, and PADs can be well described using the expression for randomly oriented species.

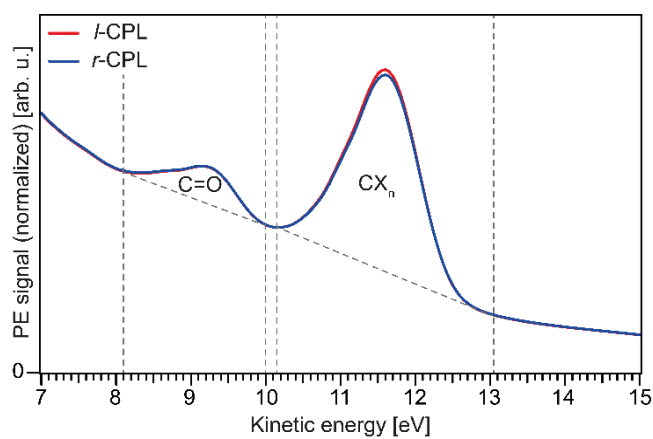

**Figure 8:** C 1s photoelectron spectra from liquid 1R-fenchone recorded at a photon energy of 302 eV with left and right circularly polarized light. The liquid fenchone temperature was stabilized to 10 °C. The large background signal (dashed line) arises from electron scattering in the liquid. Spectra were normalized to equal total area to visually suppress a small, non-essential variation of the background height when changing helicity.

An additional problem, not anticipated at the beginning of this study, is the fact that fenchone gas- and liquid-phase spectral features appear at very similar kinetic (or binding) energies. This is not the case for the core-level ionization of water, where the gas-phase O 1s peak occurs at approximately 1.8 eV lower kinetic energies than the corresponding liquid-phase O 1s peak maximum.[65] The absence of an energy shift between the liquid and the gaseous component in the case of a grounded fenchone jet reflects the nonpolar nature of this molecule and is evidence that intermolecular interactions in the liquid phase are very weak. This makes the quantification of liquid-phase PECD more complicated, as a quantitative subtraction of the overlapping gas-phase signal contribution is required. The vapor pressure of fenchone however is much lower than the one of water, with 0.3 mbar and 0.7 mbar at temperatures of 10 °C and 20 °C, which is the stabilized temperature of our sample cooling bath and the estimated maximum of the actual temperature at the glass capillary, respectively.[66] Corresponding values for water are 12.3 and 23.4 mbar. For a pure water jet, a gas-phase component as small as 5% of the total water O 1s signal can be obtained when the light focus matches the liquid-jet diameter. We aimed to find the analogous gas-phase contribution in the liquid fenchone C 1s spectrum in a separate experiment.

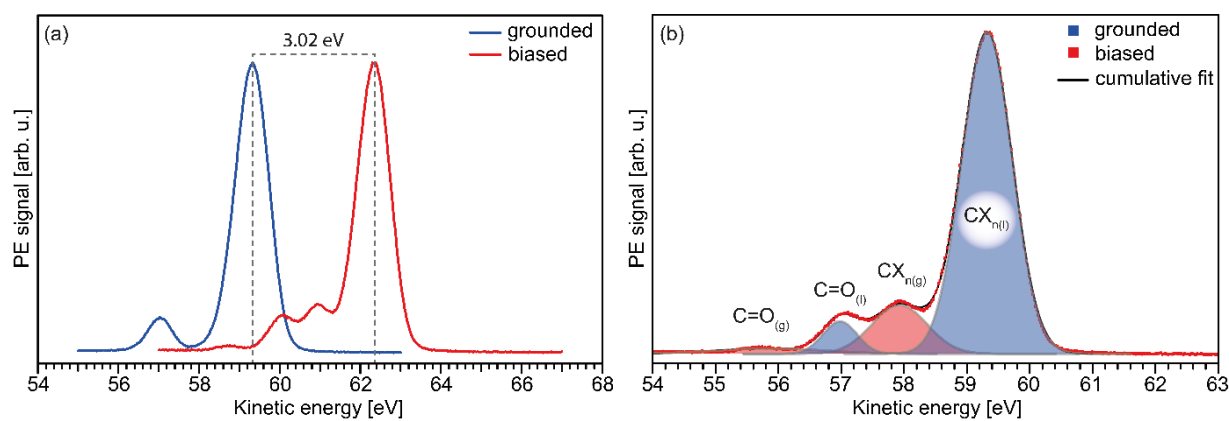

**Figure 9:** (a) C 1s photoelectron spectra from liquid 1R-fenchone recorded at a photon energy of 350 eV for a grounded liquid jet (in blue) and with an effective bias voltage of -3.02 V. Both the gas and liquid-phase peaks are affected by the bias voltage, but electrons from the gaseous species experience less overall acceleration in the weakening field gradient some distance away from the liquid jet, which resulted in a peak separation of approximately 1.3 eV between the peaks of both phases. (b) The spectrum from the biased jet has been shifted by -3.02 eV such that the liquid-phase peaks overlap for the unbiased and biased spectra. Intensities are displayed to yield same height of the main peak. See the main text for details.

Gas- and liquid-phase PE signals can be separated spectrally by applying an electric bias voltage to the liquid fenchone jet. Only the liquid-phase features experience the full energy shift from the applied potential. The gas-phase spectrum, on the other hand, experiences a smaller energy shift accompanied by spectral broadening since the strength of the accelerating field decreases with distance from the liquid jet, and electrons from the gaseous species pick up less energy on average.[23] In Figure 9a, we show a PE spectrum from 1R-fenchone, measured with a grounded liquid jet at a considerably higher photon energy of 350 eV (blue trace), using *l*-CPL. At this energy, the photoelectron spectrum is well separated from the electron scattering background tail, and the –CX$_n$ peak maximum now occurs at 59.3 eV kinetic energy, in accordance with Figure 7. The PE spectrum from the biased liquid jet (red trace in Figure 9a) exhibits four, instead of two, peaks, with the main CX$_n$ peak shifted by 3.02 eV towards higher kinetic energy. This energy shift is considerably smaller than the actual applied bias voltage of -10 V, implying resistive losses. Overall, we observe two liquid-phase peaks at 60.6 eV and 58.3 eV KE, and two less intense peaks separated from the former by 1.3 eV in KE each, *i.e.*, at 59.3 eV and 57.0 eV KE. We assign these two peaks as the gas-phase components. In Figure 9b, the biased spectrum (red dots) is shifted by 3.02 eV towards lower KE to show the liquid-phase contribution atop of the original, unbiased measurement (blue dots). We note that such rigid spectral shifts cannot be properly energy-referenced here as this would require the additional measurement of the liquid's low-energy cutoff, as discussed in Ref. 23. The observed 1.3-eV spectral separation between the liquid- and gas-phase peaks can be assigned to the difference in accelerating field strength directly at the liquid surface (for electrons originating from the liquid phase) compared to some distance away from the liquid fenchone jet (for electrons from the gas phase). The bias-induced peak separation therefore enables us to quantify the relative gas- and liquid-phase signal contributions in our PECD measurements under the given experimental conditions. For a relatively large focal spot size present in this experiment (180 µm horizontal size, 75(±10) µm vertical size) we find a gas-phase contribution of approximately 14%, inferred from a peak-area analysis. This implies that the observed PECD effect contains some contributions from the gas phase, and that the pure liquid-phase effect is likely smaller than stated in Table 1. Still, if we assume 14%-gas-phase signal contribution in our liquid-phase spectra, we arrive at a corrected $b_1^{+1}$ value of 0.015 at 302 eV for pure liquid fenchone. We believe that this estimate of the reduction in the value of $b_1^{+1}$ is rather conservative for two reasons: (1) The actual liquid fenchone measurements were done with a smaller vertical focus size of 30-40 µm (and thus a reduced gas phase fraction). (2) At similar conditions to the ones at which the gas-phase contribution for fenchone was determined, we find a gas-phase contribution of only 11% for a water jet. This last result is at odds with expectations from the relative partial gas pressures of the two solutions, but is additional evidence that our experiment provides an upper bound for the fenchone gas-phase fraction.

A quantitative analysis of the background shape and the gas-phase contribution to the measured liquid asymmetries is beyond the scope of this work, and will be presented in a separate publication. Although values given in Table 1 should grossly reflect the magnitude of the PECD effect, more accurate values remain to be determined.

In upcoming studies, we will also explore aqueous solutions, initially of chiral amino acids, where we aim to characterize the potential role of interaction with solvation-shell water molecules on PECD. Studying such aqueous solutions has the benefit that the solute contributes minimally to the gas-phase spectrum. Furthermore, future PECD studies from solution will need to address possible interfacial molecular alignment, which to some extent can be explored by comparative measurements from a planar (flat) liquid jet.

## B. Valence spectra from a liquid water flatjet obtained with He-II α radiation

Over the years, several studies reported valence photoelectron spectra from liquid water.[48, 67-69] But, to the best of our knowledge, all previously reported measurements were performed from a cylindrical liquid microjet. Indeed, cylindrical jets are easier to operate, and the smaller surface area results in a lower water-vapor background pressure; with a cylindrical jet installed we achieve $10^{-5}$ mbar in the IC of *EASI*. In conjunction with a micron-sized focus of the photon beam, approximately matching the liquid jet diameter, almost pure liquid-phase photoelectron spectra can be obtained, with gas-phase contributions as low as 5%, as mentioned earlier.[53, 70]

Yet, the curvature of a cylindrical jet implies that the measured photoelectron spectrum is an integration over all take-off angles of the photoelectrons relative to the water surface. Information related to a specific orientation of water or solvent molecules at the surface, as observed in other techniques,[71-76] is thus lost. Specifically, regarding PAD measurements from a cylindrical liquid jet, elastic electron scattering[12, 64] will contribute differently to the measured photoelectron signal intensity, which will be dependent on the unresolved take-off angle.

Being able to perform PAD measurements from a planar water surface of sub-20 μm thickness is therefore highly desirable and has motivated us to construct a flatjet system compatible with the spatial constraints of the *EASI* interaction chamber (see Section II. E). With this system, we also expect to step into novel future applications for flatjets in PES (and other X-ray spectroscopies), which are currently emerging. This would include their use in a (to be developed) liquid-jet velocity map imaging detector (LJ-VMI) or in non-linear-optical studies requiring a flat and clean solution surface, and potentially in the context of gas-liquid-surface interaction experiments, and even exploring liquid–liquid interfaces. Another, very recently discussed issue is the ability to access solution work functions[23] which would also greatly benefit from a planar surface, for the same reasons. Yet, it remains to be explored to what extent the higher gas-phase density above the flatjet surface affects the native liquid-phase PADs.

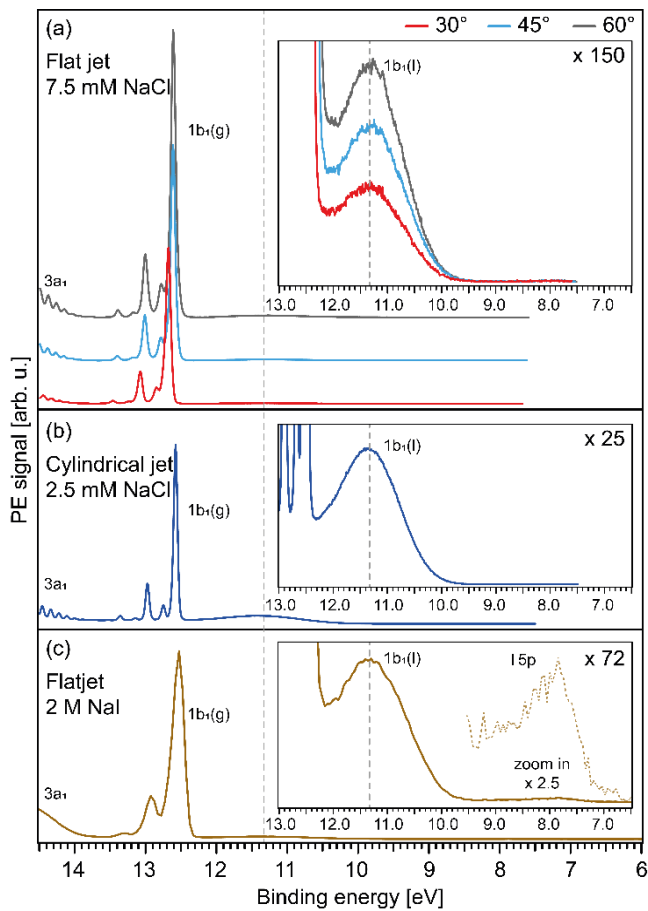

**Figure 10:** (a) $1b_1$ valence PE spectra from a liquid water (with 7.5 mM NaCl) flatjet measured with He-II α radiation (40.814 eV) at angles α = 30°, 45°, and 60°; see Figure 6 and Section II. E for experimental details. The broad small peak at 11.33 eV binding energy [23] (shown enlarged in the inset) is due to liquid water, $1b_1(l)$, and all other spectral contributions largely arise from ionization of gas-phase water, $1b_1(g)$. (b) PE spectrum as above but replacing the flatjet by a 28-μm-diameter cylindrical jet. Note that lower salt concentration, 2.5 mM NaCl, is needed to compensate for electrokinetic charging. This is due to different capillary materials and flow rates used for cylindrical *versus* flatjets. (c) Spectrum from a flatjet, as in the upper panel, at α = 45°, but from a high-concentration 2 M NaI aqueous solution, exhibiting a weak I⁻ 5p (aq.) signal near 8.2 eV binding energy. For convenience, the spectrum was shifted such that the water $1b_1(l)$ peaks of all spectra shown in the figure align at the same binding energy of 11.33 eV; note that the true water $1b_1(l)$ energy for this particular solution is 11.38 eV.[54] The small energy shift and increase of width of the gas-phase peak from the concentrated aqueous solution arises from surface charges, and hence from an electric field between flatjet surface and electron analyzer.

In Figure 10a, we show valence PE spectra from a water flatjet (see details in the caption) measured with 40.814 eV (He-II α) radiation in the *EASI* laboratory arrangement. Results are presented for three take-off angles (see Figure 6b), α = 30°, 45°, and 60°, with respect to the flatjet surface; implicitly, rotating the flatjet at fixed analyzer position also changes the angle of the incident photon beam. We display only the region of ionization of water's highest-occupied molecular orbital (HOMO), $1b_1$. The spectra are dominated by the vibrationally resolved water $1b_1(g)$ gas-phase signal contributions, originating at 12.6 eV,[77] and the $1b_1(l)$ emission from liquid water gives rise to a small signal at the lowest binding energy, shifted here to occur at 11.33 eV.[23] A zoom into the $1b_1(l)$ region is shown in the inset of Figure 10a. An observed relative increase of the liquid-phase signal of approximately 20% for larger detection angle (towards normal emission from the surface) results from the larger illuminated surface area for increasing grazing incidence angles. For comparison, Figure 10b shows a PE spectrum from a cylindrical 28-μm-diameter liquid water jet, again measured using He-II α radiation. One observes an approximately 7-8-times decrease of the liquid-to-gas signal ratio due to the – relatively – reduced evaporation from the cylindrical jet. This is accompanied by an approximately 10-fold increase of water vapor pressure when exchanging the cylindrical jet for the flatjet (without additional cryo-pumping). The overall, large gas-phase signal in all spectra of Figure 10 results from the much larger focal size of the He-II photon beam compared to the jet diameter and light foci available at synchrotron-radiation beamlines.

We note that the ability to achieve vibrational resolution of the gas-phase PE spectrum of water implies that electric fields between the liquid jet and the electron detector have been quantitatively compensated, referred to as field-free conditions. In the case of the cylindrical jet, this was accomplished by dissolving 2.5 mM NaCl in water, while for the flatjet, a higher concentration of 7.5 mM was needed due to the different capillary materials and flow rates employed; see the associated comment in Section II. E. Establishing field-free electron detection conditions from water microjets and the implications thereof are discussed in Ref. 23.

The most important conclusion from Figure 10a is that measurements of PE spectra from aqueous-solution flatjets are straightforward and can be routinely conducted even in the laboratory using a commercial (differentially pumped) He-discharge VUV source. And, more generally, routine lab-based LJ-PES measurements are very feasible and useful as we have recently demonstrated in a study of absolute ionization energies and solution work function using cylindrical jets.[23, 54] Note also that the 40.814-eV photon energy is just large enough to enable detection of the full liquid-water outer-valence band, including the $1b_2$ and $3a_1$ water orbitals. That is, the kinetic energies of the respective photoelectrons are above the aforementioned 10-12 eV limit, below which the native photoelectron peaks are highly perturbed and cannot be fully resolved. With reference to Figure 9, we note that application of a bias voltage equally works for a liquid-water flatjet, and should enable spectral separation of liquid-phase from gas-phase features. In fact, if the bias voltage is large enough an essentially gas-phase-free PE spectrum can be obtained, so far demonstrated only for cylindrical jets though.[23, 54] Hence (disturbing) large gas-phase spectral contributions in Figure 10 can be expected to be elegantly and almost quantitatively removed. We conclude by presenting a valence PE spectrum from a flatjet from an aqueous solution in Figure 10c, exemplified here for 2 M NaI measured with the same 40.814 eV. This particular spectrum was measured for eight minutes which is sufficient to detect the iodide $I^-$ $5p_{1/2}$ and $I^-$ $5p_{3/2}$ signal.[23] Systematic and high-signal-to-noise PE spectra from flatjets of different solutions will be presented in an upcoming work.

## IV. SUMMARY

We have presented a unique experimental setup, *EASI*, with all major components, that enables PECD and regular PAD measurements, associated with the chiral $b_1$ and non-chiral $b_2$ anisotropy parameters, respectively, from liquid microjets of (aqueous) solutions. *EASI*'s principal configurations – one for PECD and three for regular PAD measurements – and how transformation between those configurations is accomplished time-effectively and with rather little effort, have been described in detail. Regarding *EASI*'s performance, near-ionization-threshold C 1s photoelectron spectra from 1S- and 1R- gas-phase and liquid fenchone for different helicities of circularly polarized soft-X-rays have been presented. This aspect of our work shows the feasibility of liquid-jet PECD studies, also highlighting the difficulty of quantifying PECD for a non-polar liquid. Our results encourage studies from chiral molecules in aqueous solutions. With respect to laboratory experiments, conducted in conjunction with a commercial VUV light source, we have presented photoelectron spectra from a planar microjet (flatjet) for the first time, exemplified here for neat liquid water and NaI aqueous solution. Planar jets will play a crucial role in developing the field of PADs from liquid surfaces – but in conjunction with small focal sizes of XUV to soft X-ray beamlines – and also in designing novel efficient detectors, simultaneously collecting signal over a large electron emission angle.

## ACKNOWLEDGMENTS

The authors would like to thank Laurent Nahon, Ivan Powis, Stephan Thürmer, and Iain Wilkinson for the many stimulating discussions particularly in the planning and commissioning phase of *EASI*. We also thank Stephan Thürmer and Iain Wilkinson for critical reading of and comments on this manuscript, as well as their assistance with the collection and processing of the fenchone data. We further acknowledge discussion with Oleg Kornilov and his team regarding the construction of the jet recirculation system. B.W. acknowledges funding from the European Research Council (ERC) under the European Union's Horizon 2020 research and innovation program under grant agreement No. GAP 883759 - AQUACHIRAL. S.M., U.H., and B.W. acknowledge support by the Deutsche Forschungsgemeinschaft (Wi 1327/5-1). F.T. and B.W. acknowledge support by the MaxWater initiative of the Max-Planck-Gesellschaft. All authors acknowledge DESY (Hamburg, Germany), a member of the Helmholtz Association HGF, for the provision of experimental facilities. Parts of this research were carried out at PETRA III and we would like to thank Moritz Hoesch in particular, as well as the whole beamline staff, the PETRA III chemistry laboratory and crane operators for assistance in using the P04 soft X-ray beamline. Beamtime was allocated for proposal II-20180012 (LTP). Additionally, we thank the FHI workshop for supporting us with the technical realization and especially for last-minute modifications of the setup. We acknowledge discussion with Vincent Lehane (Scienta

Omicron, Uppsala, Sweden) and for providing a custom-made lens table that enables the quantitative detection of the full low-energy tail of the photoelectrons from liquid jets.## DATA AVAILABILITY

Data of relevance to this study have been deposited at the following DOI: 10.5281/zenodo.5730418.

## CONFLICTS OF INTEREST

There are no conflicts to declare.